\documentclass[runningheads,a4paper]{llncs}

\usepackage{amssymb}
\usepackage{graphicx}
\usepackage{todonotes}
\usepackage{booktabs}
\usepackage[inline]{enumitem}
\newlist{inlinelist}{enumerate*}{1}
\setlist*[inlinelist,1]{label=\roman*),itemjoin={{, }},itemjoin*={{, and }}}

\usepackage{url}
\newcommand{\keywords}[1]{\par\addvspace\baselineskip
\noindent\keywordname\enspace\ignorespaces#1}
\usepackage{appendix}

\begin{document}

\mainmatter  

\title{A Study on Token Pruning for ColBERT}

\titlerunning{A Study on Token Pruning for ColBERT}

%
%
\author{Carlos Lassance$^\dagger$
\and Maroua Maachou$^\dagger$ \and Joohee Park$^\ddagger$\and Stéphane Clinchant$^\dagger$} 
\authorrunning{Lassance et al.}

\institute{$^\dagger$Naver Labs Europe, $^\ddagger$ Naver Corp\\
$^\dagger$first.lastname@naverlabs.com, $^\ddagger$ first.lastname@navercorp.com   \\
}

%
%

\maketitle

\begin{abstract}
The ColBERT model has recently been proposed as an effective BERT based ranker. By adopting a late interaction mechanism, a major advantage of ColBERT is that document representations can be precomputed in advance. However, the big downside of the model is the index size, which scales linearly with the number of tokens in the collection. In this paper, we study various designs for ColBERT models in order to attack this problem. While compression techniques have been explored to reduce the index size, in this paper we study token pruning techniques for ColBERT. We compare simple heuristics, as well as a single layer of attention mechanism to select the tokens to keep at indexing time. Our experiments show that ColBERT indexes can be pruned up to 30\% on the MS MARCO passage collection without a significant drop in performance. Finally, we experiment on MS MARCO documents, which reveal several challenges for such mechanism.
\keywords{information retrieval, token pruning, BERT, ColBERT}
\end{abstract}

\section{Introduction}

In recent Information Retrieval (IR) systems, Pretrained Language Models (PLM)~\cite{lin2020IRBertReview} have taken the state of the art by storm. Two main families of PLM retrieval methods have been developed: \begin{enumerate}
    \item Representation-based~\cite{reimers2019sentence}, where both query and documents are encoded separately into a single representation and scoring is performed via distance between representations;
    \item Interaction-based~\cite{passage_ranking}, where a query-document pair is treated jointly by a neural network to generate a score.
\end{enumerate}  
On one hand, the former is very efficient as representations of documents can be indexed and only the query has to be computed during inference time. On the other hand, the latter has better performance as it is able to perform a more thorough scoring between queries and documents. In order to bridge the gap between these two families, the ColBERT~\cite{colbert} method indexes a representation per token, which allows to precompute document representations and part of the capability of an interaction model (each contextualized token of the query interacts with each precomputed contextualized token of the document). However, the ColBERT advantage comes with an important cost on the index size, since every token (rather than a pooled version of a document) needs to be indexed.

In this work we investigate this method, by looking into two characteristics of the representations: normalization and query expansion. We then focus on the index size by limiting the amount of tokens to be saved in each document using 4 methods based on \begin{inlinelist} \item position of the token \item Inverse Document Frequency (IDF) of the token \item special tokens \item attention mechanism of the last layer \end{inlinelist}. To empirically evaluate our method, we perform our investigation under the most common benchmark for neural IR (MS MARCO on both passage and document tasks), showing that we are able to greatly improve efficiency (in terms of index size and complexity) while still maintaining acceptable efficacy (in terms of MRR and Recall). Finally, we note that many approaches have addressed this problem from the token compression side, by reducing the representation size of each individual token, which could be ultimately combined with our approach.

\section{Related work}

\paragraph{Efficient Pretrained LMs:}
In NLP, there has been a lot of work seeking to improve the efficiency of
pretrained LMs. For instance, quantization and distillation have been extensively studied in this context \cite{sanh2019distilbert}.
Closest to our work, Power-BERT \cite{power-bert} and length adaptive transformers \cite{kim-cho-2021-length} have been proposed to reduce the number of FLOPS operations, by eliminating tokens in the PLM layers. 

\paragraph{Index pruning in IR:} Pruning indexes is a traditional method in Information Retrieval to reduce latency and memory requirements and has been studied thoroughly in many contributions (e.g~\cite{zobel2006inverted} and ~\cite{carmel2001static}), as far as we are aware this is the first application to PLM token-level indexing. 

\paragraph{Improving ColBERT Efficiency}
One way to mitigate the problem of the increased index size of ColBERT is to reduce the dimensionality and apply quantization to the document tokens. Note that this is already done by default as the output of the PLM (most of the time of 768 dimensions) is projected into a smaller space (128 dimensions). In the original paper, the authors show very good performance by both reducing the dimensionality and quantizing the tokens~\cite{colbert}. Recently, a binarization technique~\cite{yamada2021efficient} has been proposed for information retrieval and preliminary code and experiments in the official ColBERT repository~\footnote{\url{https://github.com/stanford-futuredata/ColBERT/tree/binarization}} show that it should be possible to binarize the ColBERT tokens. 

The present work is orthogonal to the previously presented research direction, in that we aim to reduce the index size by removing tokens from the index, instead of reducing their size. We consider the combination of those research directions as necessary to improve ColBERT models, but leave it as future work. Lastly, ColBERT has been extended to perform pseudo relevance feedback in \cite{colbert_prf} and query pruning has been studied to improve latency in \cite{colbert_query_pruning}.

\section{Methodology and ColBERT}

The ColBERT model is based on a transformer~\cite{bert} encoder for documents and queries. Each item $Q$ or $D$ is encoded into a representation $\mathbf{R}$ which is a matrix of dimensions $(t,d)$ where $t$ is the amount of tokens in the item and $d$ is the encoding dimension of each token. The score of a document ($D$) for a given query ($Q$) is given by the following formula:

\begin{equation}
    s(D,Q) = \sum_{j=0}^{|t_Q|} \max_{i=0}^{|t_D|} (\mathbf{R}_D \cdot \mathbf{R}_Q)_{i,j}\;.
\end{equation}

This late-interaction model improves computation efficiency compared to a full-interaction model, because  representations $\mathbf R_D$ for all documents ($D \in \mathcal{D}$) can be precomputed. In this way, during inference only the query representation needs to be computed. However, this incurs a very large representation size by document ($dt$) which can quickly become intractable when the amount of documents increases. As those are the only factors impacting index size, there are two solutions: \begin{inlinelist} \item reduce the amount of dimensions per token \item reduce the amount of tokens per document \end{inlinelist}. While many solutions exist for the first case, we are not aware of any previous work for the second one.

\paragraph{Normalization and query expansion: }

In an effort to ensure that the scores of each query and document are not dependent on their length, each token is normalized in the l2-hyper-sphere. Also queries are expanded so that they have the same length (32 tokens). We show via experiments that this is not actually needed for all PLM backbones (c.f. Section~\ref{results-colbert}), which reduces the inference cost significantly.

\subsection{Limiting the amount of tokens}
\label{methods}
We investigate 4 diverse methods for limiting the amount of tokens that are stored for each document. These methods are integrated \textbf{during the training} of our ColBERT models\footnote{We tested both during and just after training and noticed that integrating this during training improved the results.}. In other words, we add a pooling layer on top of Colbert
to select a maximum of $k$ tokens per document and then use the Colbert scoring equation restricted to the k selected tokens.

\paragraph{\textbf{First} $k$ tokens:}

One very simple, but also very strong, baseline is to keep only the $k$ first tokens of a document. Indeed such a baseline takes advantage of the inherent bias of the MS MARCO dataset where the first tokens are considered to be the most important~\cite{hofstatter2021mitigating}. This is the only method where we do not remove punctuation tokens.

\paragraph{Top $k$ \textbf{IDF} tokens:}

Another possibility is to choose the rarest tokens of a document, in other words, the ones with the highest Inverse Document Frequency (IDF). This should allow us to keep only the most defining words of a document.

\paragraph{\textbf{$k$-Unused Tokens}}

We also test the possibility of adding $k$ special tokens ('unused' terms from the BERT vocabulary) to the start of the document and keep only those tokens. In this way, the kept tokens are always the same for all documents and always in the same positions, which increases the consistency for the model which we posit leads to easier optimization. However this approach has some drawbacks: \begin{inlinelist} \item increased computation time as it forces documents to have at least $k$ tokens \item possibly missing context from long documents as  truncation is done with less ``real-word'' tokens.\end{inlinelist}

\paragraph{\textbf{Attention score:}}

We propose to use the \emph{last layer of attention of the PLM} as a way to detect the most important $k < t$ tokens of the document. Recall that the attention of a document ($A_D$) is a three-dimensional tensor ($h,t,t'$), where the first dimension is the number of attention heads (number of different "attentions") and the last two dimensions represent the amount of attention a token $t$ "pays" to token $t'$, which is normalized so that for each token $t$ the sum of attentions $t'$ is 1. We propose an importance score per token which is the sum of attention payed to the token over all heads:

\begin{equation}
\label{eq:attention score}
    i(t)_D = \sum_{i=0}^{h} \sum_{j=0}^{|t_D|} (A_D)_{i,t,j}
\end{equation}

So that at the end we only keep $k < t$ tokens, based on the top-$k$ scores. Note that this attention score is computed over the \emph{last layer of the document encoding}, and not in the interaction between documents and queries. In other words it is \emph{independent} from queries.

\section{Experiments}

\paragraph{Experimental setup: }

In this work, we train our ColBERT~\cite{colbert} models using the MINILM-12-384~\cite{wang2020minilm}\footnote{available at \url{https://huggingface.co/microsoft/MiniLM-L12-H384-uncased}} as the PLM base. Models are trained during 150k steps, with a linear learning rate from 0 to the final learning rate (2e-5) warmup of 10k steps and linear learning rate annealing (2e-5 to 0) for the remaining 140k steps. We use the original triplet files provided by the MS MARCO organizers. We evaluate our models in the full ranking scenario, using a brute force implementation, which differs from the original ColBERT paper, which uses approximate nearest neighbors for that scenario. By not using ANN search, we avoid the need of ANN-hyperparameter tuning and/or the risk of unfair comparisons. Nevertheless, we have experimented with ANN and results are very close to the ones presented in the following paragraphs.

\paragraph{Investigating default ColBERT: }
\label{results-colbert}
First, we investigate the default ColBERT model and verify if the normalization and query expansion operations are helpful for all PLM. Results for this ablation are presented in Table~\ref{table:ablation}. We observe that when using MINILM instead of BERT as our PLM there is no need for normalization and expansion. Therefore we can skip these steps which reduces the complexity of the search (search complexity depends linearly to the amount of tokens in the query). For the rest of this work, we only use models without these operations.

\begin{table}[ht]
\centering
\caption{Ablation of token normalization and query expansion on the MINILM ColBERT model.}
\label{table:ablation}
\begin{tabular}{cc|ccc}
\toprule
Normalization & Expansion & MRR@10 & Recall@1000 \\
\midrule
X             & X         & 0.362  & 96.3\%      \\
X             &           & 0.055  & 41.2\%      \\
              & X         & 0.363  & 97.0\%      \\
              &           & \textbf{0.365}  & \textbf{97.1\%}     \\
\bottomrule
\end{tabular}
\end{table}

\paragraph{Reducing index size on passage dataset: }

We now evaluate the ability of the 4 proposed token pruning methods (c.f. Section~\ref{methods}). Results are displayed in Table~\ref{table:passage}. We notice that when keeping the top $k=50$, almost all the tested methods allow us to reduce the amount of tokens by 30\% while keeping similar acceptable performance (less than 0.01 MRR@10 drop, less than 1\% Recall@1000 drop). On the other hand, when we drop to $k=10$ there is a noticeable drop in MRR@10 performance for all methods, with \emph{Unused Tokens} getting the best results, which shows that this technique is very useful for small $k$ but not for larger ones. Note that further tests in other datasets could be useful to verify not only this difference between \emph{Unused Tokens} and the other methods, but also of the \emph{First} method that uses the inherent bias of MS MARCO to the start of the document~\cite{hofstatter2021mitigating}. Note that \% of tokens may vary due to: the size of passages (\emph{$k$=50}); the use or not of punctuation (\emph{First}); repeated scores (\emph{attention} and \emph{IDF}) or because a method increases the original passage length (\emph{Unused Tokens}).

\begin{table}[ht]
\centering
\caption{Results for different methods and $k$ tokens to keep on MS MARCO. We use MRR@10 for the passage dataset and MRR@100 for the document one. Index size consider 2kb per vector (128 dimensions in fp16).}
\label{table:passage}
\begin{tabular}{r|cc|cc}
\toprule
Method                          & \% tokens & Size (Gb) & MRR  & Recall@1000 \\
\midrule
Baseline                        & 100\%  & 142   & 0.365     & 97.1\%      \\
\midrule
\multicolumn{5}{c}{Passage (k=50)}                                                       \\
\midrule
First                           & 72.6\%  & 103  & 0.357  & 96.7\%            \\
IDF                             & 71.9\%  & 102  & 0.355  & 96.7\%            \\
Unused Tokens                           & 74.1\% & 101   & 0.342  &  96.2\%  \\     
Attention Score                 & 71.1\% & 105  & \textbf{0.358}  &            96.7\% \\ 
\midrule
\multicolumn{5}{c}{Passage (k=10)}                                                       \\
\midrule
First                           & 14.8\% & 21   & 0.302  & 93.3\%      \\
IDF                             & 15.3\% & 22  & 0.290  & 91.0\%      \\
Unused Tokens                             & 14.8\% & 21  & \textbf{0.314}    & \textbf{94.0\%}      \\ 
Attention Score                 & 14.8\% & 21   & 0.281  & 91.9\%      \\
\midrule
\multicolumn{5}{c}{Document (k=50)} \\
\midrule
Baseline                        & 100\% & 290   & 0.380   & 95.6\%      \\
\midrule
First                           & 13.1\% & 38  & 0.347   & 91.6\%      \\
IDF                             & 13.5\% & 39  & 0.225   & 81.2\%      \\
Unused Tokens                   & 13.1\% & 38  & \textbf{0.354}   & \textbf{93.1\%}      \\
Attention Score                 & 13.1\% & 38  & 0.306   & 90.9\%      \\
\bottomrule
\end{tabular}
\end{table}

\paragraph{Reducing index size on document dataset: }

In the case of the document task, the mean amount of tokens per document is higher than in the case of passages. Due to this disparity we test only the case of $k=50$, which allows us to cut 85\% of the document tokens. While $k=50$ provides a noticeable reduction of index size, it still means that the amount of data to keep for every document is large ($50d$). As it was the case with the smaller token size on the passage task, both \emph{Unused Tokens} and \emph{First} present the best results.

\paragraph{Results using ANN}

For completeness we also run retrieval of passages following the approximate nearest neighbor scheme described in the original ColBERT paper~\cite{colbert} and available at the original source code: \url{https://github.com/stanford-futuredata/ColBERT/}. For the first stage token retrieval we use a probing of the 128 closests clusters and perform full retrieval on all documents that had tokens in the top 8k of each query token. Results are available in Table~\ref{appendix-table:passage}.

\begin{table}[ht]
\centering
\caption{Results for different methods and $k$ tokens to keep on MSMarco-passage using ANN retrieval}
\label{appendix-table:passage}
\begin{tabular}{r|c|cc}
\toprule
Method                          & \% tokens & MRR@10 &  Recall@1000 \\
\midrule
Baseline                        & 100\%     & 0.365   & 96.5\%      \\
\midrule
\multicolumn{4}{c}{k=50}                                                       \\
\midrule
First                           & 72.6\%    & 0.357   & 96.0\%           \\
IDF                             & 71.9\%    & 0.355  & 96.0\%            \\
Attention Score                 & 71.1\%    & \textbf{0.358}     & \textbf{96.2\%}            \\
\midrule
\multicolumn{4}{c}{k=10}                                                       \\
\midrule
First                           & 14.8\%    & \textbf{0.302}           &    \textbf{93.2\%}         \\
IDF                             & 15.3\%    & 0.290  &    90.9\%         \\
Attention Score                 & 14.8\%    & 0.281  &    92.1\%         \\
\bottomrule
\end{tabular}
\end{table}

\section{Result analysis}

\paragraph{Analysis of indexed documents: }
We analyzed the tokens selected by the attention mechanism on the document set. One problem we observed is about repetitions: it selects of the same token at different position. For instance, we give an example below of the selected tokens for one document to demonstrate the repetition and the stemming effect (dog vs dogs):

{ \it \small
    ['[CLS]', 'dog', 'nasal', 'poly', 'removal', 'dog', 'nasal', 'poly', 'removal', 'poly', 'grow', 'body', 'parts', 'dogs', 'nasal', 'poly', 'dog', 'being', 'dog', 'nasal', 'poly', 'dog', 'dog', 'nasal', 'poly', 'removal', 'nasal', 'poly', 'dogs', 'poly', 'dog', 'dogs', 'nasal', 'poly', 'dogs', 'hide', '’', 'dogs', 'dog', 'nasal', 'growth', 'nasal', 'poly', 'dogs', '’', 'dogs', 'nose', 'dogs', 'nose', '[SEP]']}

These results indicates that token selection methods should take into account either the repetition problem or model the diversity of the selected tokens, which we leave for future work.


\paragraph{ColBERT relevance: }

Note that while ColBERT was the state of the art for MS MARCO full ranking during its release, it has been overtook by dense~\cite{tas-balanced,cocodenser} and sparse~\cite{splade}  with better training procedure, namely: \begin{inlinelist} \item hard-negative mining~\cite{tas-balanced,cocodenser,splade} \item distillation~\cite{tas-balanced,splade} \item pretraining~\cite{cocodenser} \end{inlinelist}. When this work was conducted, these techniques were not tested for the ColBERT model but were recently shown to be beneficial in \cite{colbert_v2}. Finally, ColBERT has demonstrated interesting properties such as better zero-shot performance~\cite{beir} and the combination with traditional IR techniques such as PRF~\cite{colbert_prf}.

\section{Conclusion}

In this work we investigate the ColBERT model and test different methods to reduce its late-interaction complexity and its index size. We first verify that for some PLM (namely MINILM) we do not need to perform normalization of tokens and query augmentation. We also note that some very simple methods (such as keeping the first $k$ tokens) can allow us to remove 30\% of the tokens from the passage index without incurring in any consequent performance loss. On the other hand the MS MARCO document collection reveals challenges for such mechanisms, where even pruning 90\% of tokens may not be enough. The combination of such token pruning methods with already studied embedding compression methods could lead to further improvement of ColBERT-based IR systems.

\bibliographystyle{abbrv}

\bibliography{main}   

\end{document}